\documentclass[runningheads]{llncs}
\usepackage{graphicx}
\usepackage{ulem}
\usepackage{soul}
\usepackage{multirow}
\usepackage[colorlinks,linkcolor=blue]{hyperref}

\usepackage{tikz}
\usepackage{comment}
\usepackage{amsmath,amssymb} 
\usepackage{color}

\begin{document}
\pagestyle{headings}
\mainmatter
\def\ECCVSubNumber{108}  

\title{Efficient Image Super-Resolution \\
	Using Pixel Attention} 

\titlerunning{Efficient Image Super-Resolution Using Pixel Attention}
\author{Hengyuan Zhao\inst{1,2} \and
Xiangtao Kong\inst{1,2,3} \and
Jingwen He\inst{1,2} \and \\
Yu Qiao\inst{1,2} \and
Chao Dong\inst{1,2}}
\authorrunning{Hengyuan Zhao et al. }
%
\institute{ShenZhen Key Lab of Computer Vision and Pattern Recognition, SIAT-SenseTime Joint Lab, Shenzhen Institutes of Advanced Technology, Chinese Academy of Sciences\\ \and
	SIAT Branch,  Shenzhen Institute of Artificial Intelligence and Robotics for Society\\ \and
	University of Chinese Academy of Sciences \\
	\email{\{hy.zhao1, xt.kong, jw.he, yu.qiao, chao.dong\}@siat.ac.cn}}
\maketitle

\begin{abstract}

  This work aims at designing a lightweight convolutional neural network for image super resolution (SR). With simplicity bare in mind, we construct a pretty concise and effective network with a newly proposed pixel attention scheme. Pixel attention (PA) is similar as channel attention and spatial attention in formulation. The difference is that PA produces 3D attention maps instead of a 1D attention vector or a 2D map. This attention scheme introduces fewer additional parameters but generates better SR results. On the basis of PA, we propose two building blocks for the main branch and the reconstruction branch, respectively. The first one — SC-PA block has the same structure as the Self-Calibrated convolution but with our PA layer. This block is much more efficient than conventional residual/dense blocks, for its two-branch architecture and attention scheme. While the second one — U-PA block combines the nearest-neighbor upsampling, convolution and PA layers. It improves the final reconstruction quality with little parameter cost. Our final model — PAN could achieve similar performance as the lightweight networks — SRResNet and CARN, but with only 272K parameters (17.92\% of SRResNet and 17.09\% of CARN). The effectiveness of each proposed component is also validated by ablation study. The code is available at \href{https://github.com/zhaohengyuan1/PAN}{https://github.com/zhaohengyuan1/PAN}.
  
\end{abstract}

  \section{Introduction}
  Image super resolution is a long-standing low-level computer vision problem, which predicts a high-resolution image from a low-resolution observation. In recent years, deep-learning-based methods \cite{dong2015image} have dominated this field, and consistently improved the performance. Despite of the fast development, the huge and increasing computation cost has largely restricted their application in real-world usages, such as real-time zooming and interactive editing. To address this issue, the AIM 2020 \cite{ignatov2020aim_bokeh,ignatov2020aim_ISP,fuoli2020aim_VXSR,ntavelis2020aim_inpainting,elhelou2020aim_relighting,wei2020aim_realSR,son2020aim_VTSR} held the ``Efficient Super Resolution" challenge \cite{zhang2020aim_efficientSR}, which asked the participants to use fewer computation cost to achieve the same performance as a standard baseline – SRResNet \cite{ledig2017photo}. This challenge could significantly promote the development of light-weight networks. With the goal of minimizing parameters, we propose an extremely simple yet effective model - PAN, which has fewest parameters among all participants. We will introduce our method in this paper. The review of previous studies can be found in the Related Work section. 
  Our main contribution is called pixel attention (PA), which is inspired by channel attention (CA) \cite{hu2018squeeze} and spatial attention (SA) \cite{woo2018cbam}. These attention schemes are popular as they can effectively improve the feature representation capacity by a ``second order" feature multiplication. By applying a more powerful feature propagation strategy, the network could achieve higher performance with the same computation load (e,g., RCAN \cite{zhang2018image}, CARN \cite{ahn2018fast}, PANet\cite{mei2020pyramid}). This is a promising direction for network compression. Specifically, as shown in Fig.\ref{fig:5}, channel attention pools the previous features to a vector by spatial global pooling, while spatial attention pools the features to a single feature map by channel-wise pooling. We find that these schemes are less effective in SR task, which requires pixel-level evaluation. On the other hand, simply removing the pooling operation could significantly improves the performance. As the features are multiplied in a pixel-wise manor (see Fig.\ref{fig:5}), we call this modified attention scheme as pixel attention, and our network as pixel attention network (PAN).
  
  \begin{figure}[t]
    \centering
    \includegraphics[width=10cm]{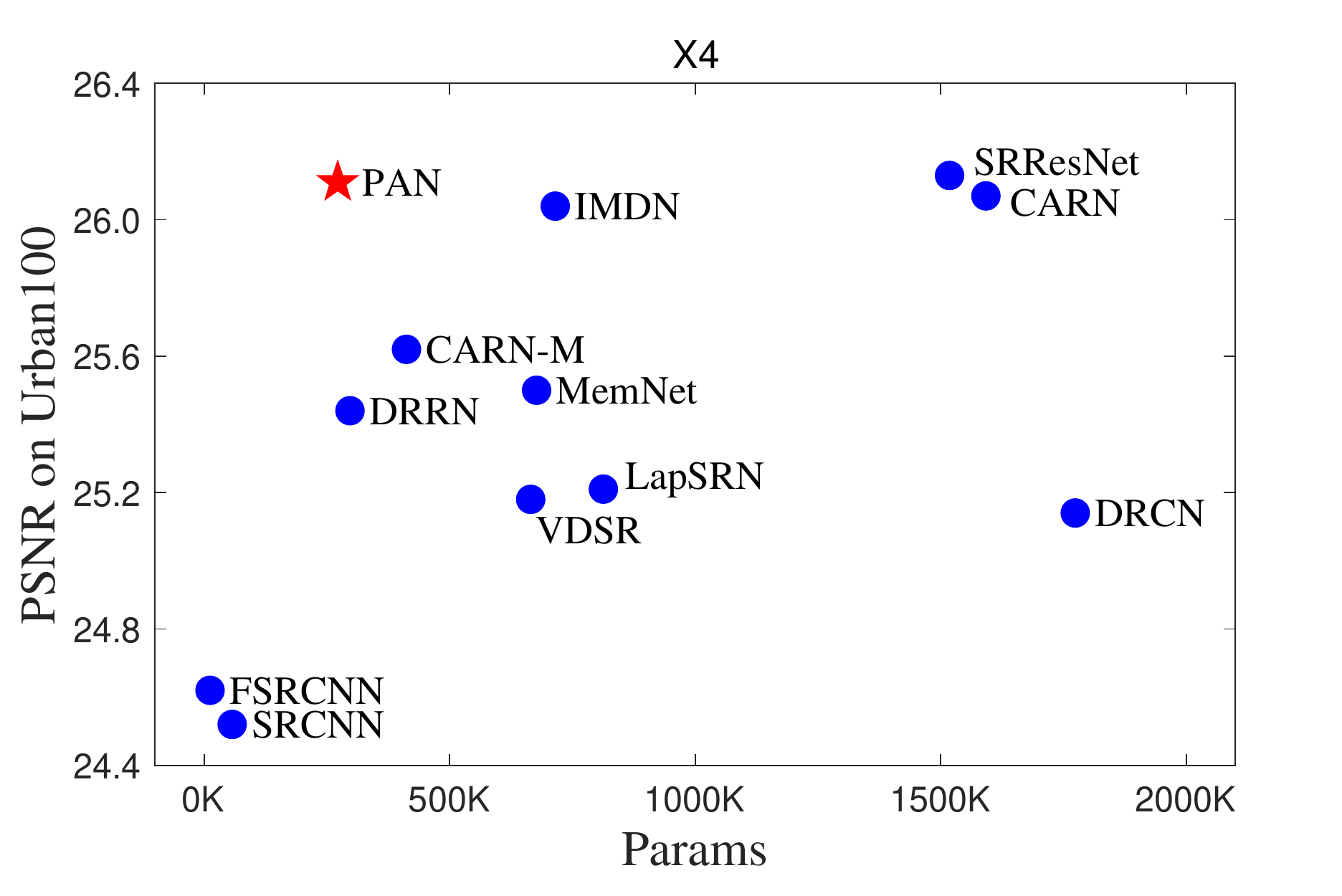}
    \caption{Performance and Parameters comparison between our PAN and other state-of-the-art lightweight networks on Urban100 dataset for upscaling factor $\times4$.}
    \label{fig:4}
  \end{figure}

  We equip PA in two building blocks, which forms the whole network. The first block is the basic block in the main branch, called Self-Calibrated block with Pixel Attention (SC-PA). Its main structure is the same as the recent Self-Calibrated Convolutions \cite{liu2020improving}. As shown in Fig.\ref{fig:1}, the convolutions are divided into two portions: the upper one is responsible for higher-level feature manipulation, while the other one is to maintain the original information. We adopt PA in the upper one, and standard convolutions in the other one. The SC-PA block enjoys a very simple structure without complex connections and up/down sampling operations, which are not friendly for hardware acceleration.
  
  The second block is the Upsampling block with Pixel Attention  (U-PA), which is in the reconstruction stage. It is based on the observation that previous SR networks mainly adopt similar structures in the reconstruction branch, i.e., deconvolution/pixel-shuffle layers and regular convolutions. Little efforts are devoted in this area. However, this structure is found to be redundant and less effective. To further improve the efficiency, we introduce PA between the convolution layers and use nearest neighbor (NN) upsampling to further reduce parameters. Thus the U-PA block consists of NN, convolution, PA and convolution layers, as shown in Fig.\ref{fig:1}. 
  
  The overall framework of PAN is pretty simple in architecture, yet is much more effective than previous models. In AIM 2020 challenge, we are ranked 4th in overall ranking, as we are inferior in the number of convolutions and activations. However, our entry contains the fewest parameters -- only 272K, which is 161K fewer than the 1st, and 415K fewer than the 2nd. From Fig.~\ref{fig:4}, we observe that PAN achieves a better trade-off between the reconstruction performance and the model size. To further realize the potential of our method, we try to expand PAN to a larger size. However, it will become hard to train the larger PAN without adding connections in the network. But if we add connections (e.g. dense connections) in PAN, it will dramatically increase the computation, which is not consistent with our goal -- efficient SR. We believe that PA is an useful and independent component, which could also benefit other computer vision tasks.

  \textbf{Contributions.} The main contributions of this work are threefold:

  1.We propose a simple and fundamental attention scheme -- pixel attention (PA), which is demonstrated effective in lightweight SR networks.

  2.We integrate pixel attention and Self-Calibrated convolution \cite{liu2020improving} in a new building block -- SC-PA, which is efficient and constructive.

  3.We employ pixel attention in the reconstruction branch and propose a U-PA block. Note that few studies have investigated attention schemes after upsampling.

  \section{Related Work}
 \subsection{Efficient CNN for SR}
 
 Recently, many deep neural networks \cite{zhang2018residual,zhang2018image,dai2019second,Lim_2017_CVPR_Workshops,zhang2019ranksrgan,he2019multi} have been introduced to improve the reconstruction results. However, the huge amount of parameters and the expensive computational cost limit their practice in real applications \cite{ahn2018fast}.
 
 To save the computation, Dong et al. \cite{dong2016accelerating} directly use the original LR images as input instead of the pre-upsampled ones. This strategy has been widely used in SISR models \cite{dong2016accelerating,shi2016real,wang2018esrgan}. Besides, group convolution \cite{bevilacqua2012low,matsui2017sketch,He_2019_CVPR}, depth-wise separable convolutions \cite{timofte2017ntire,huang2015single}, and self-calibrated convolution \cite{liu2020improving} have been proposed to accelerate the deep models. Some of these modules have been utilized in SR and shows effectiveness \cite{ahn2018fast,wang2020deep}. CARN-M \cite{ahn2018fast} uses group convolution for efficient SR and obtains comparable results against computational complexity models. IMDN \cite{hui2019lightweight} extracts hierarchical features step-by-step by split operations, and then aggregates them by simply using a 1×1 convolution. It won the first place at Contrained Super-Resolution Challenge in AIM 2019 \cite{zhang2019aim}. In this work, we employ the self-calibrated convolution scheme \cite{liu2020improving} in our PAN networks for efficient SR.
 
 \subsection{Attention Scheme}
 
 Attention mechanism has demonstrated great superiority in improving  the performance of deep models for computer vision tasks. SE-Net \cite{hu2018squeeze} is the first attention method to learn channel information and achieves state-of-the-art performance. Roy et al. \cite{roy2018concurrent} uses a $1 \times 1$ convolution layer to generate spatial attention features. BAM \cite{park2018bam} decomposes 3D attention map inference into channel and spatial attention map. CBAM \cite{woo2018cbam} computes spatial attention using a 2D convolution layer of kernel size $k \times k$, then combines it with channel attention to generate 3D attention map. ECA \cite{wang2020eca} employes global average pooling (GAP) to generate channel weights by performing a fast 1D convolution of size $k$. Zhang et al. \cite{zhang2018image} proposed the residual channel attention network (RCAN) by introducing the channel attention mechanism into a modified residual block for SR. The channel attention mechanism uses global average pooling to extract channel statistics. It can rescale channel-wise features by considering interdependencies among channels to help train a very deep network. Dai et al. \cite{dai2019second} proposed the second-order attention network (SAN) by using second-order feature statistics for more discriminative representations.

 Obviously, most of the above methods focus on developing complex attention modules to gain better performance. Different from them, our PA aims at learning effective pixel attention with lower computation complexity and generates 3D attention features with a $1\times1$ convolution layer.
 
 \subsection{Reconstruction Methods in SR Networks}
 
 Instead of adopting interpolation based upsampling methods at the beginning of network\cite{dong2015image,kim2016accurate,tai2017image}, the learning-based reconstruction methods such as pixel-shuffle \cite{shi2016real} generally implement upsampling in the final stage of the network. But in recent works, interpolation based upsampling methods can also be employed in the end of network to obtain good performance \cite{wang2018esrgan}. Therefore, the reconstruction module now basically consists of upsampling (interpolation based or learning based) and convolutional layers. Our reconstruction method in PAN adopts interpolation based -- nearest neighbor upsampling and convolution layers.
 
 Besides, previous works have shown that attention mechanism can effectively improve the performance in SR tasks but few researchers investigate it in reconstruction stage. Therefore, in this work, our U-PA block based on attention mechanism is adopted in reconstruction stage for better reconstruction. 
  
  \section{Proposed Method}

  \subsection{Network Architecture}

  \begin{figure}[t]
    \centering
    \includegraphics[width=\linewidth]{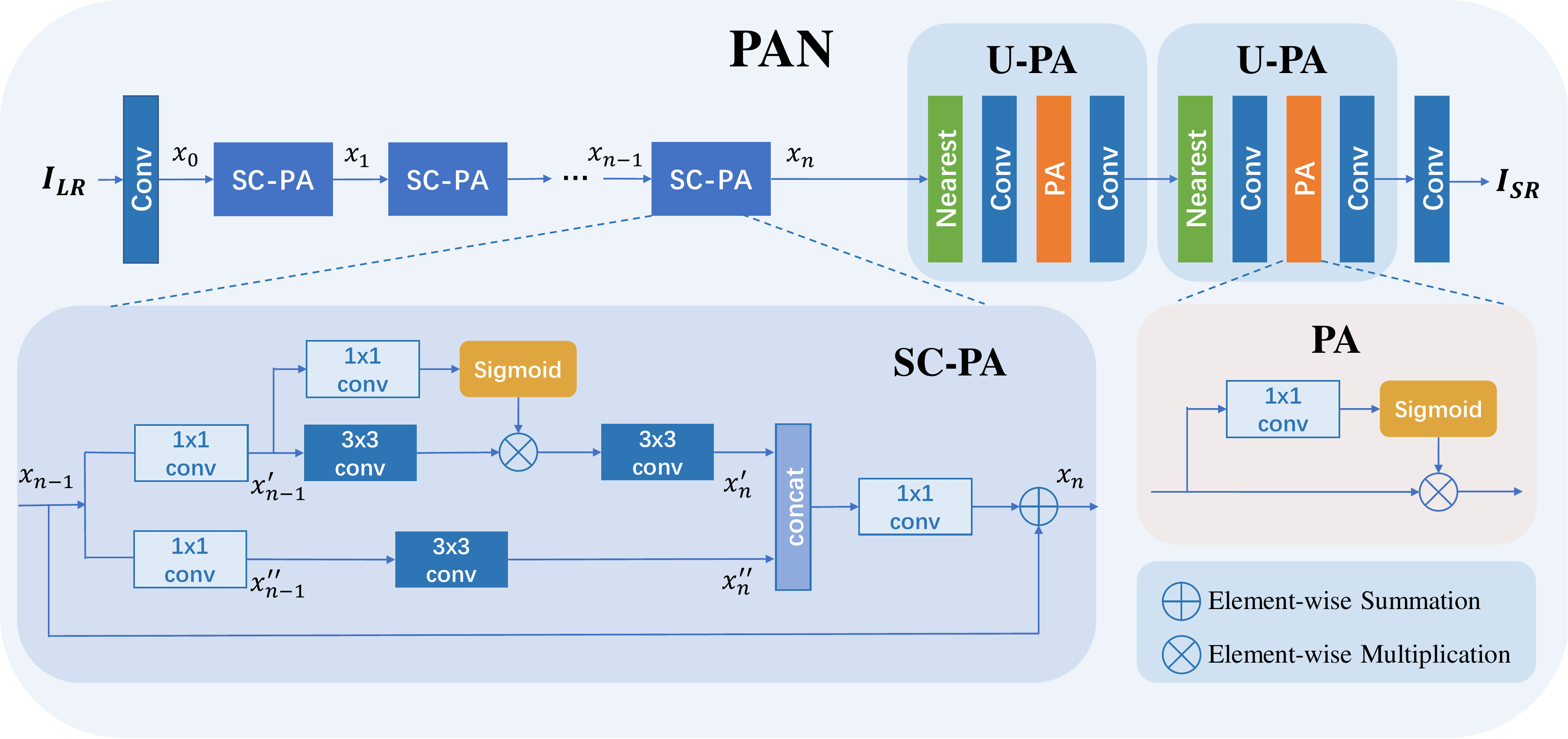}
    \caption{Network architecture of the proposed PAN.}
    \label{fig:1}
  \end{figure}
  
  As shown in Fig.\ref{fig:1}, the network architecture of our
  PAN, consists of three modules, namely the
  feature extraction (FE) module, the nonlinear mapping module
  with stacked SC-PAs, and the reconstruction module with U-PA blocks.

  The LR images are first fed to the FE module that contains a
  convolution layer for shallow feature extraction. The FE module
  can be formulated as
  \begin{align}
   x_0 = f_{ext}(I_{LR}),
  \end{align}
  where $f_{ext}(\cdot)$ denotes a convolution layer with a 3$\times$3 kernel to extract features from the input LR image $I_{LR}$, and $x_0$ is the extracted feature maps. It is worth noting that only one convolution layer is used here for lightweight design.

  Then, we use the non-linear mapping module that consists of several stacked SC-PAs to generate new powerful feature representations. We denote the proposed SC-PA as $f_{SCPA}(\cdot)$ given by
  \begin{align}
    x_n = f_{SCPA}^{n}(f_{SCPA}^{n-1}(...f_{SCPA}^{0}(x_0)...)),
  \end{align}
  where $x_n$ is the output feature map of the $n$th SC-PA.

  At last, we utilize the reconstruction module that contains two U-PA blocks and a convolution layer to upsample the features to the HR size. In addition, we add a global connection path $f_{UP}$, in which a bilinear interpolation is performed on the input $I_{LR}$. Finally, we obtain:
  \begin{align}
    I_{SR} = f_{rec}(x_n)+f_{up}(I_{LR}),
  \end{align}
  where $f_{rec}(\cdot)$ is the reconstruction module, and $I_{SR}$ is the final result of the network. 
  
  \subsection{Pixel Attention Scheme}

  \begin{figure}[t]
    \centering
    \includegraphics[width=11cm]{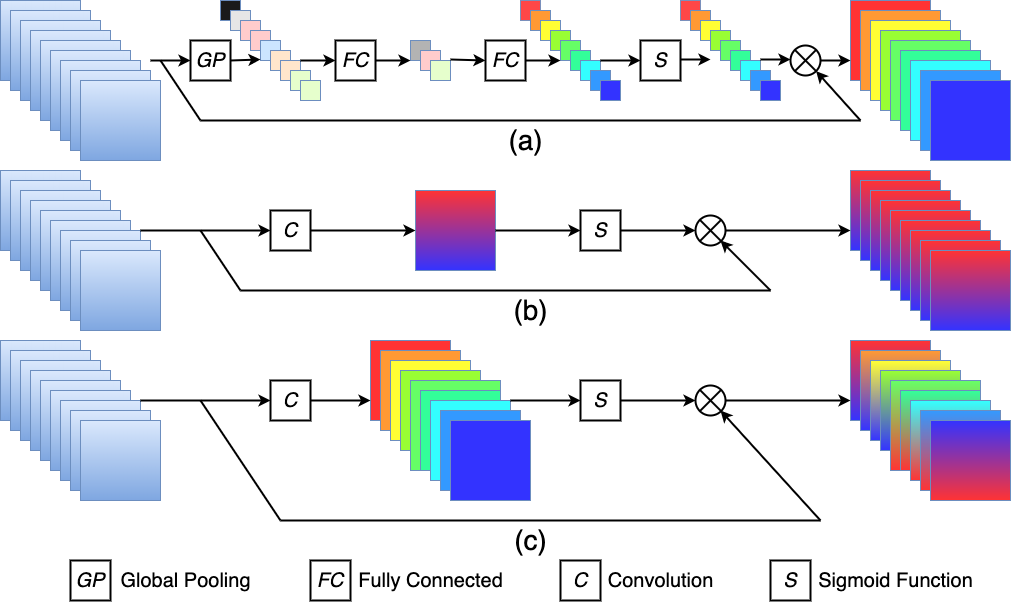}
    \caption{(a) CA: Channel Attention; (b) SA: Spatial Attention; (c) PA: Pixel Attention.}
    \label{fig:5}
  \end{figure}

  First, we revisit channel attention \cite{hu2018squeeze} and spatial attention\cite{woo2018cbam}. As shown in Fig.\ref{fig:5}, channel attention aims to obtain a 1D ($C \times 1 \times1$) attention feature vector, while spatial attention obtains a 2D ($1 \times H \times W$) attention map. Note that $C$ is the number of channels, $H$ and $W$ are the height and width of the features, respectively. Different from them, our pixel attention is able to generate a 3D  ($C \times H \times W$) matrix as the attention features. In other word, pixel attention generates attention coefficients for all pixels of the feature map. As shown in Fig.\ref{fig:2}, pixel attention only uses a $1\times1$ convolution layer and a sigmoid function to obtain the attention maps which will then be multiplied with the input features.

  we denote the input and output feature map as $x_{k-1}$ and $x_k$, respectively. The PA layer can be computed as
  \begin{align}
    x_k = f_{PA}(x_{k-1}) \cdot x_{k-1},
  \end{align}
  where $f_{PA}(\cdot)$ is a $1\times1$ convolution layer followed by a sigmoid function.
  
  \subsection{SC-PA Block}
  
  The nonlinear mapping module contains several stacked Self-Calibrated convolution with Pixel Attention (SC-PA) blocks. 
  Here, we define $x_{n-1}$ and $x_{n}$ as the input and output of the $n$th SC-PA block, respectively. As shown in Fig.\ref{fig:1}, similarly as SCNet\cite{liu2020improving}, SC-PA block contains two branches, where each branch has a $1\times1$ convolution layer at the beginning, which is called $f_{split}(\cdot)$. Given input feature $x_{n-1}$, we have:
  \begin{align}
    {x_{n-1}}' = f_{split}'(x_{n-1}),\\
    {x_{n-1}}'' = f_{split}''(x_{n-1}),
  \end{align}
  where ${x_{n-1}}'$ and ${x_{n-1}}''$ only have  half of the channel number of ${x_{n-1}}$.

  The upper branch also contains two $3\times3$ convolution layers, where the first one is equipped with a pixel attention module. This branch transforms $x_{n-1}'$ to $x_{n}'$. We only use a single $3\times3$ convolution layer to generate $x_{n}''$ for the purpose of maintaining the original information. Finally, $x_{n}'$ and $x_{n}''$ are concatenated and then passed to a $1\times1$ convolution layer to generate $x_{n}$. In order to accelerate training, shortcut is used to produce the final output feature $x_{n}$. This block is inspired by SCNet \cite{liu2020improving}. The main difference is that we employ our PA scheme to replace the pooling and upsampling layer in SCNet \cite{liu2020improving}.
  
  \subsection{U-PA Block}
  
  Besides the nonlinear mapping module, pixel attention is also adopted in the final reconstruction module. As shown in Fig.\ref{fig:2}, the U-PA block consists of a nearest neighbor (NN) upsampling layer and a PA layer between two convolution layers. Note that in previous SR networks, a reconstruction module is basically comprised of upsampling and convolution layers. Moreover, few researchers have investigated the attention mechanism in the upsampling stage. Therefore, in this work, we adopt PA layer in the reconstruction module. Experiments show that introducing PA could significantly improve the final performance with little parameter cost. Besides, we also use the nearest-neighbor interpolation layer as the upsampling layer to further save parameters.

  \subsection{Discussion}
  The proposed PAN is specially designed for efficient SR, thus is very concise in network architecture. The building blocks – SC-PA and U-PA are also simple and easy to implement. Nevertheless, when expanding this network to a larger scale, i.e., $\textgreater$50 blocks, the current structure will face the problem of training difficulty. Then we need to add other techniques, like dense connections, to allow successful training. As this is not the focus of this paper, we do not investigate these additional strategies for very deep networks. There is also another limitation for the proposed PA scheme.
  
  We experimentally find that PA is especially useful for small networks. But the effectiveness decreases with the increase of network scales. This is mainly because that PA can improve the expression capacity of convolutions, which could be very important for lightweight networks. In contrast, large-scale networks are highly redundant and their convolutions are not fully utilized, thus the improvement will mainly comes from a better training strategy. We have shown this trend in ablation study. 
  
  \section{Experiments}
  In this section, we systematically compare our PAN with state-of-the-art SISR algorithms on five commonly used benchmark datasets. Besides, we conduct ablation study to validate the effectiveness of each proposed component.
 
  \subsection{Datasets and Metrics}
  We use DIV2K and Flickr2K datasets as our training datasets. The LR images are obtained by the bicubic downsampling of HR images. During the testing stage, five standard benchmark datasets, Set5 \cite{bevilacqua2012low}, Set14 \cite{yang2010image}, B100 \cite{martin2001database}, Urban100 \cite{huang2015single}, Manga109 \cite{matsui2017sketch}, are used for evaluation. The widely used peak signal to noise ratio (PSNR) and the structural similarity index (SSIM) on the Y channel are used as the evaluation metrics.
  
  \subsection{Implementation Details}
  
  During training, we use DIV2K and Flickr2K to train our PAN. For training SRResNet-PA and RCAN-PA, we only use DIV2K dataset. Data augmentation is also performed on the training set by random rotations of 90$^{\circ}$, 180$^{\circ}$, 270$^{\circ}$ and horizontal flips. The HR patch size is set to 256$\times$256, while the minibatch size is 32. L1 loss function \cite{wang2004image} is adopted with Adam optimizer \cite{kingma2014adam} for model training. The cosine annealing learning scheme rather than the multi-step scheme is adopted since it has a faster training speed. The initial maximum learning rate is set to $1e-3$ and the minimum learning rate is set to $1e-7$. The period of cosine is $250k$ iterations. The proposed algorithm is implemented under the PyTorch framework \cite{paszke2017automatic} on a computer with an NVIDIA GTX 1080Ti GPU.
  
  \subsection{Comparison with SRResNet and CARN}
  PAN is dedicated for the efficient SR challenge\cite{zhang2020aim_efficientSR}. According to the requirements, our aim is to achieve at least the same performance as the SRResNet \cite{ledig2017photo} in the provided validation dataset with lower computational cost. 
  In this section, we mainly compare with SRResNet on the aforementioned five standard benchmark datasets for super resolution $\times2$,  $\times3$, and  $\times4$.
  Besides, we select another state-of-the-art network -- CARN \cite{ahn2018fast}, which is specially designed to be efficient and lightweight. 
  
  From Table~\ref{tab:5}, it is obviously observed that our proposed PAN outperforms CARN on five benchmark datasets for all upscaling factors $\times2$,  $\times3$, and  $\times4$. Note that the number of parameters in PAN only accounts for less than 1/5 of CARN \cite{ahn2018fast}. As for comparison with SRResNet \cite{ledig2017photo}, our proposed PAN could obtain higher PSNR on Manga109 and B100 dataset, but yeilds inferior results on the other three datasets.
  Specifically, for task $\times$4, the number of parameters in PAN is only 17.92\% of SRResNet \cite{ledig2017photo} and 17.09\% of CARN \cite{ahn2018fast}, respectively.

\begin{table}[htbp]
  \centering
  \caption{Comparison of SRResNet, CARN and PAN for upscaling factors $\times$2, $\times$3, and $\times$4. Red/Blue text: best/second-best.}
  \label{tab:5}
  \renewcommand{\arraystretch}{1.2}
 \begin{tabular}{ccccccccc}
	\hline
	Scale &
	Method &
	Params &
	Mult-Adds &
	\begin{tabular}[c]{@{}c@{}}Set5\\ PSNR\end{tabular} &
	\begin{tabular}[c]{@{}c@{}}Set14\\ PSNR\end{tabular} &
	\begin{tabular}[c]{@{}c@{}}B100\\ PSNR\end{tabular} &
	\begin{tabular}[c]{@{}c@{}}Urban100\\ PSNR\end{tabular} &
	\begin{tabular}[c]{@{}c@{}}Manga109\\ PSNR\end{tabular} \\ \hline
	& CARN      & 1,592K                        & 222.8G        & 37.76 & 33.52 & 32.09 & 31.92 & {\color[HTML]{0000FE}38.36} \\
	& SRResNet  & 1,370K &341.7G           & {\color[HTML]{FE0000}38.05} & {\color[HTML]{FE0000}33.64} & {\color[HTML]{FE0000}32.22} & {\color[HTML]{FE0000}32.23} & 38.05 \\
	\multirow{-3}{*}{$\times$2} & PAN(Ours) & 261K                         & 70.5G & {\color[HTML]{0000FE}38.00} & {\color[HTML]{0000FE}33.59} & {\color[HTML]{0000FE}32.18} & {\color[HTML]{0000FE}32.01} & {\color[HTML]{FE0000}38.70} \\ \hline
	& CARN      & 1,592K                        & 118.8G          & 34.29 & 30.29 & 29.06 & 28.06 & 33.50 \\
	& SRResNet  & 1,554K                        &   190.2G        & {\color[HTML]{FE0000}34.41} & {\color[HTML]{FE0000}30.36} & {\color[HTML]{FE0000}29.11} & {\color[HTML]{FE0000}28.20} & {\color[HTML]{0000FE}33.54} \\
	\multirow{-3}{*}{$\times$3} & PAN(Ours) & 261K                         & 39.0G & {\color[HTML]{0000FE}34.40} & {\color[HTML]{0000FE}30.36} & {\color[HTML]{0000FE}29.11} & {\color[HTML]{0000FE}28.11} & {\color[HTML]{FE0000}33.61} \\ \hline
	& CARN      & 1,592K                        &    90.9G       & 32.13 & 28.60 & 27.58 & 26.07 & 30.47 \\
	& SRResNet  & 1,518K                        & 146.1G    & {\color[HTML]{FE0000}32.17} & {\color[HTML]{FE0000}28.61} & {\color[HTML]{0000FE}27.59} & {\color[HTML]{FE0000}26.12} & {\color[HTML]{0000FE}30.48} \\
	\multirow{-3}{*}{$\times$4} & PAN(Ours) & 272K                         & 28.2G    & {\color[HTML]{0000FE}32.13} & {\color[HTML]{0000FE}28.61} & {\color[HTML]{FE0000}27.59} & {\color[HTML]{0000FE}26.11} & {\color[HTML]{FE0000}30.51} \\ \hline
\end{tabular}
  \end{table}
  
  \subsection{Ablation Study}
  \begin{figure}[t]
    \centering
    \includegraphics[width=10cm]{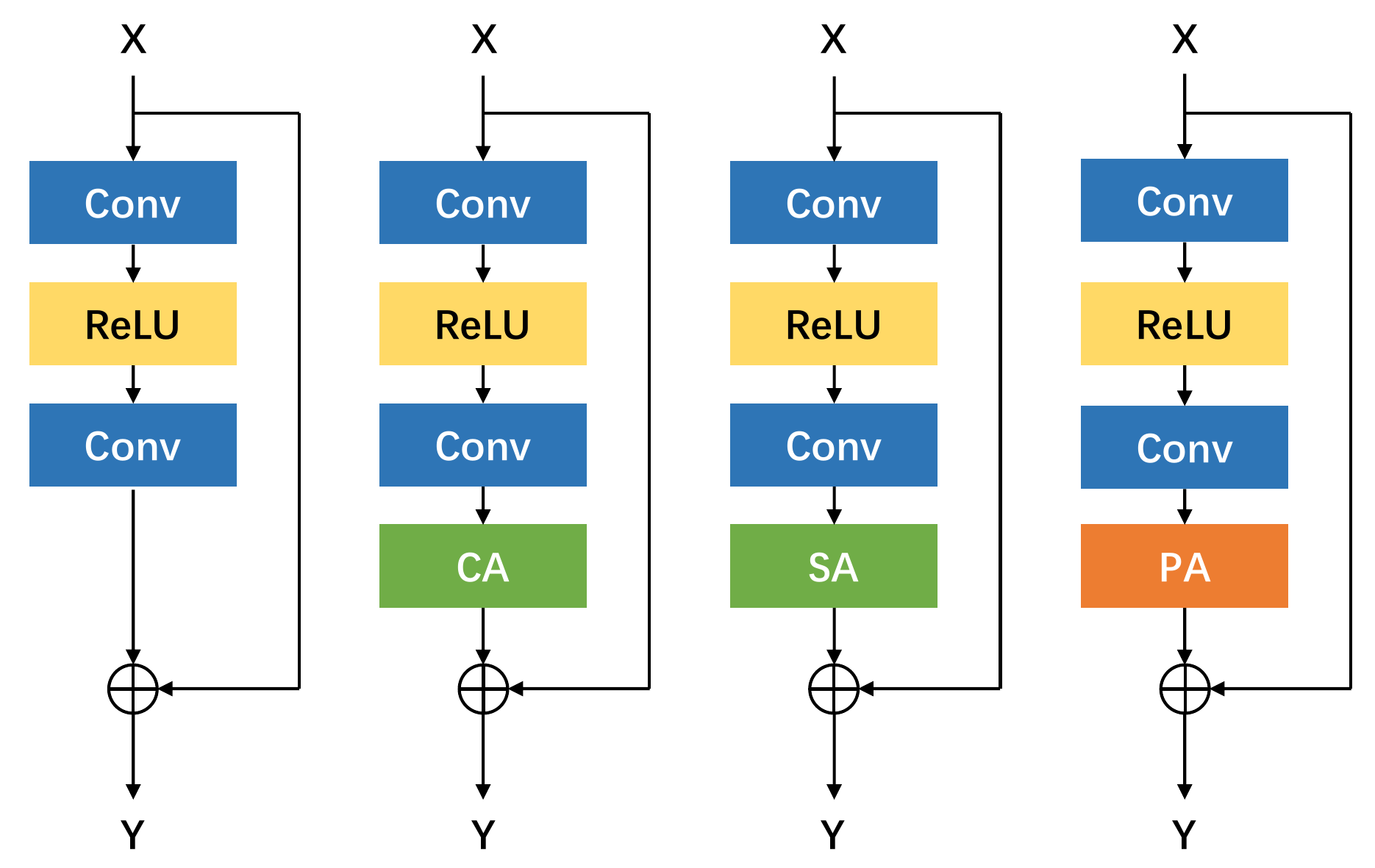}
    \caption{(a) RB: basic residual block; (b) RB-CA: basic residual block with channel attention; (c) RB-SA: basic residual block with spatial attention; (d) RB-PA: basic residual attention with pixel attention.}
    \label{fig:2}
  \end{figure}
  
  \subsubsection{Comparison of different attention schemes.}
  To demonstrate the effectiveness of our PA layer, we use PAN as the basic network, and then replace the 16 SC-PA blocks with 8 residual blocks (RB), 8 residual blocks with channel attention (RB-CA), 8 residual blocks with spatial attention (RB-SA) and 8 residual blocks with pixel attention (RB-PA), respectively. As shown in Fig.\ref{fig:2}, the attention module is inserted after the second convolution in the original residual block, which is consistent with other attention schemes.
  
  \setlength{\tabcolsep}{4pt}
  \renewcommand{\arraystretch}{1.2}
  \begin{table}[htbp]
    \begin{center}
      \caption{Comparison of the number of parameters, Mult-Adds and mean values of PSNR obtained by Basic RB, RB-CA, RB-SA and RB-PA on five datasets for upscaling factor $\times$4. We record the results in $5 \times 10^{5}$ iterations.}
      \label{tab:1}
      \begin{tabular}{llll}
        \hline
        Attention Type & Params & Mult-Adds  & PSNR               \\ \hline
        RB             & 272,009 & 28.16G & 27.94dB        \\
        RB-CA           & 285,379 & 28.16G & 27.97dB({\color[HTML]{FE0000}+0.03dB})          \\
        RB-SA           & 272,427 & 28.18G & 27.93dB({\color[HTML]{00FE00}$-$0.01dB})         \\
        RB-PA           & 285,219 & 28.90G	 & 28.03dB({\color[HTML]{FE0000}+0.09dB}) \\ \hline
  
        \end{tabular}
    \end{center}
    \end{table}
  \setlength{\tabcolsep}{1.4pt}
  
  In Table~\ref{tab:1}, we compare the number of parameters, Multi-Adds, and the performance in PSNR for all methods. Note that all results are the mean values of PSNR calculated by 328 images on 5 benchmark datasets. Mult-Adds is computed by assuming that the resolution of HR image is 720p. 
  It is observed that RB-CA and RB-PA could improve the PSNR by 0.03dB and 0.09dB, respectively, while RB-SA is slightly worse than RB. This indicates that pixel attention is more effective than channel attention and spatial attention.
  
  \subsubsection{The effectiveness of Self-Calibrated (SC) block.}
  We validate the effectiveness of the Self-Calibrated (SC) block by comparing SC-PA with RB-PA. Here, we use RB as a baseline. From Table~\ref{tab:2}, we find that both RB-PA and SC-PA could improve the PSNR on the basis of RB. Furthermore, SC-PA outperforms RB-PA by 0.12dB and it only requires 410 (1/30 of RB-PA) additional parameters compared with RB.
  This demonstrates that the Self-Calibrated (SC) block is able to achieve more significant improvement compared with the traditional residual blocks.

  \setlength{\tabcolsep}{4pt}
  \renewcommand{\arraystretch}{1.2}
    \begin{table}[htbp]
      \centering
      \caption{Comparison of the number of parameters and mean values of PSNR obtained by Basic RB, RB-PA and SC-PA on five datasets for upscaling factor $\times$4. We record the results in $5 \times 10^{5}$ iterations.}
      \label{tab:2}
      \begin{tabular}{llll}
      \hline
      Basic Unit   & RB      & RB-PA                         & SC-PA             \\ \hline
      Params Diff. & 0       & +13,210                        & +410              \\
      Mult-Adds    & 28.16G  & 28.90G & 28.16G            \\
      PSNR         & 27.94dB & 28.03dB ({\color[HTML]{FE0000}+0.09dB})            & 28.15dB ({\color[HTML]{FE0000}+0.21dB}) \\ \hline
      \end{tabular}
      \end{table}
  \setlength{\tabcolsep}{1.4pt}
  
  \subsubsection{The effectiveness of PA.}
  Here, we show the importance of pixel attention (PA) in the Self-Calibrated (SC) block and Upsampling (U) block. 
  For baseline, we remove all PA in our proposed PAN. As shown in Table~\ref{tab:3}, we observe that adding pixel attention (PA) in Self-Calibrated (SC) block and Upsampling (U) block could achieve improvement by 0.06 dB.

    \setlength{\tabcolsep}{4pt}
    \renewcommand{\arraystretch}{1.2}
    \begin{table}[htbp]
      \centering
      \caption{The effectiveness of PA in SC-PA and U-PA blocks measured on the five benchmark datasets for upscaling factor $\times4$. We record the PSNR in $5 \times 10^{5}$ iterations.}
      \label{tab:3}
      \begin{tabular}{c|cccc}
      \hline
      PA in SC-PA  & $\times$       & $\surd$                        & $\times$       & $\surd$       \\
      PA in U-PA   & $\times$       & $\times$                        & $\surd$       & $\surd$       \\ \hline
      Params & 264,499  & 271,219                   & 265,699  & 272,419  \\
      PSNR   & 28.09dB & 28.12dB                  & 28.08dB & 28.15dB \\ \hline
      \end{tabular}
      \end{table}
  \setlength{\tabcolsep}{1.4pt}

  \subsubsection{The influence of model size.}
  We also investigate the effectiveness of pixel attention (PA) in networks with different model sizes. 
 For comparison, we select two networks, SRResNet and RCAN, whose number of parameters are 1,518K and 15,592K, respectively. Then, we add PA to both of the two networks and name them as SRResNet-PA and RCAN-PA, respectively. Besides, we remove PA from PAN and name the new network as PAN-SC-U. 
 Since training larger network requires more time, here we record the results in $1 \times 10^{6}$ iterations.
 As we can see from Table~\ref{tab:6}, PA brings increase in PSNR (0.11dB) on the basis of PAN-SC-U. However, it seems that PA could degrade the performance of the larger networks (SRResNet-PA and RCAN-PA). For instance, RCAN-PA is worse than RCAN with a 0.06dB drop in PSNR. 
The experimental results shows that PA is more effective in lightweight models.

\begin{table}[htbp]
	\centering
	\caption{Quantitative results of state-of-the-art SR methods for all upscaling factors $\times2$, $\times3$, and $\times4$. Red/Blue text: best/second-best among all methods except RDN, RCAN, and SAN. Overstriking: our methods.}
	\label{tab:4}
	\renewcommand{\arraystretch}{0.95}
	\begin{tabular}{|l|c|c|c|c|c|c|}
		\hline
		Method &
		Params &
		\begin{tabular}[c]{@{}c@{}}Set5\\ PSNR/SSIM\end{tabular} &
		\begin{tabular}[c]{@{}c@{}}Set14\\ PSNR/SSIM\end{tabular} &
		\begin{tabular}[c]{@{}c@{}}B100\\ PSNR/SSIM\end{tabular} &
		\begin{tabular}[c]{@{}c@{}}Urban100\\ PSNR/SSIM\end{tabular} &
		\begin{tabular}[c]{@{}c@{}}Manga109\\ PSNR/SSIM\end{tabular} \\ \hline
		Scale     & \multicolumn{6}{c|}{$\times$2}                                                           \\ \hline
		SRCNN     & 57K     & 36.66/0.9542 & 32.45/0.9067 & 31.36/0.8879 & 29.50/0.8946 & 35.60/0.9663 \\
		FSRCNN &
		{\color[HTML]{000000} 13K} &
		37.00/0.9558 &
		\multicolumn{1}{l|}{32.63/0.9088} &
		31.53/0.8920 &
		\multicolumn{1}{l|}{29.88/0.9020} &
		\multicolumn{1}{l|}{36.67/0.9710} \\
		VDSR      & 666K   & 37.53/0.9587 & 33.03/0.9124 & 31.90/0.8960 & 30.76/0.9140 & 37.22/0.9750 \\
		DRCN      & 1,774K  & 37.63/0.9588 & 33.04/0.9118 & 31.85/0.8942 & 30.75/0.9133 & 37.55/0.9732 \\
		LapSRN    & 251K   & 37.52/0.9591 & 32.99/0.9124 & 31.80/0.8952 & 30.41/0.9103 & 37.27/0.9740 \\
		DRRN      & 298K   & 37.74/0.9591 & 33.23/0.9136 & 32.05/0.8973 & 31.23/0.9188 & 37.88/0.9749 \\
		MemNet    & 678K   & 37.78/0.9597 & 33.28/0.9142 & 32.08/0.8978 & 31.31/0.9195 & 37.72/0.9740 \\
		CARN      & 1,592K  & 37.76/0.9590 & 33.52/0.9166 & 32.09/0.8978 & 31.92/0.9256 & 38.36/0.9765 \\
		SRResNet  & 1,370K  & {\color[HTML]{FE0000}38.05}/{\color[HTML]{FE0000}0.9607} & {\color[HTML]{FE0000}33.64}/{\color[HTML]{0000FE}0.9178} & {\color[HTML]{FE0000}32.22}/{\color[HTML]{FE0000}0.9002} & {\color[HTML]{FE0000}32.23}/{\color[HTML]{FE0000}0.9295} & 38.05/0.9607 \\
		IMDN      & 694K   & 38.00/0.9605 & {\color[HTML]{0000FE}33.63}/0.9177 & {\color[HTML]{0000FE}32.19}/0.8996 & {\color[HTML]{0000FE}32.17}/{\color[HTML]{0000FE}0.9283} & {\color[HTML]{FE0000}38.88}/{\color[HTML]{FE0000}0.9774} \\
		\textbf{PAN}       & \textbf{261K}   & {\color[HTML]{0000FE}38.00}/{\color[HTML]{0000FE}0.9605} & 33.59/{\color[HTML]{FE0000}0.9181} & 32.18/{\color[HTML]{0000FE}0.8997} & 32.01/0.9273 & {\color[HTML]{0000FE}38.70}/{\color[HTML]{0000FE}0.9773} \\ \hline
		RDN       & 22,123K & 38.24/0.9614 & 34.01/0.9212 & 32.34/0.9017 & 32.89/0.9353 & 39.18/0.9780 \\
		RCAN      & 15,444K & 38.27/0.9614 & 34.12/0.9216 & 32.41/0.9027 & 33.34/0.9384 & 39.44/0.9786 \\
		SAN      & 15,674K & 38.31/0.9620& 34.07/0.9213& 32.42/0.9028 &33.10/0.9370& 39.32/.09792 \\
		\hline
		Scale     & \multicolumn{6}{c|}{$\times$3}                                                           \\ \hline
		SRCNN     & 57K     & 32.75/0.9090 & 29.30/0.8215 & 28.41/0.7863 & 26.24/0.7989 & 30.48/0.9117 \\
		FSRCNN    & 13K    & 33.18/0.9140 & 29.37/0.8240 & 28.53/0.7910 & 26.43/0.8080 & 31.10/0.9210 \\
		VDSR      & 666K   & 33.66/0.9213 & 29.77/0.8314 & 28.82/0.7976 & 27.14/0.8279 & 32.01/0.9340 \\
		DRCN      & 1,774K  & 33.82/0.9226 & 29.76/0.8311 & 28.80/0.7963 & 27.15/0.8276 & 32.24/0.9343 \\
		LapSRN    & 502K   & 33.81/0.9220 & 29.79/0.8325 & 28.82/0.7980 & 27.07/0.8275 & 32.21/0.9350 \\
		DRRN      & 298K   & 34.03/0.9244 & 29.96/0.8349 & 28.95/0.8004 & 27.53/0.8378 & 32.71/0.9379 \\
		MemNet    & 678K   & 34.09/0.9248 & 30.00/0.8350 & 28.96/0.8001 & 27.56/0.8376 & 32.51/0.9369 \\
		CARN      & 1,592K  & 34.29/0.9255 & 30.29/0.8407 & 29.06/0.8034 & 28.06/0.8493 & 33.50/0.9440 \\
		SRResNet  & 1,554K  & {\color[HTML]{FE0000}34.41}/{\color[HTML]{FE0000}0.9274} & {\color[HTML]{FE0000}30.36}/{\color[HTML]{FE0000}0.8427} & {\color[HTML]{FE0000}29.11}/{\color[HTML]{FE0000}0.8055} & {\color[HTML]{FE0000}28.20}/{\color[HTML]{FE0000}0.8535} & 33.54/{\color[HTML]{0000FE}0.9448} \\
		IMDN      & 703K   & 34.36/0.9270 & 30.32/0.8417 & 29.09/0.8046 & {\color[HTML]{0000FE}28.17}/{\color[HTML]{0000FE}0.8519} & {\color[HTML]{0000FE}33.61}/0.9445 \\
		\textbf{PAN}       & \textbf{261K}   & {\color[HTML]{0000FE}34.40}/{\color[HTML]{0000FE}0.9271} & {\color[HTML]{0000FE}30.36}/{\color[HTML]{0000FE}0.8423} & {\color[HTML]{0000FE}29.11}/{\color[HTML]{0000FE}0.8050} & 28.11/0.8511 & {\color[HTML]{FE0000}33.61}/{\color[HTML]{FE0000}0.9448} \\ \hline
		RDN       & 22,308K & 34.71/0.9296 & 30.57/0.8468 & 29.26/0.8093 & 28.80/0.8653 & 34.13/0.9484 \\
		RCAN      & 15,629K & 34.74/0.9299 & 30.65/0.8482 & 29.32/0.8111 & 29.09/0.8702 & 34.44/0.9499 \\ 
		SAN      & 15,859K &34.75/0.9300& 30.59/0.8476 &29.33/0.8112& 28.93/0.8671& 34.30/0.9494 \\
		\hline
		Scale     & \multicolumn{6}{c|}{$\times$4}                                                           \\ \hline
		SRCNN      & 57K                        & 30.48/0.8628 & 27.49/0.7503 & 26.90/0.7101 & 24.52/0.7221 & 27.66/0.8505 \\
		FSRCNN     & 12K & 30.71/0.8657 & 27.59/0.7535 & 26.98/0.7105 & 24.62/0.7280 & 27.90/0.8517 \\
		VDSR       & 665K                       & 31.35/0.8838 & 28.01/0.7674 & 27.29/0.7251 & 25.18/0.7524 & 28.83/0.8809 \\
		DRCN       & 1,774K                      & 31.53/0.8854 & 28.02/0.7670 & 27.23/0.7233 & 25.14/0.7510 & 28.98/0.8816 \\
		LapSRN     & 813K                       & 31.54/0.8850 & 29.19/0.7720 & 27.32/0.7280 & 25.21/0.7560 & 29.09/0.8845 \\
		DRRN       & 297K                       & 31.68/0.8888 & 28.21/0.7720 & 27.38/0.7284 & 25.44/0.7638 & 29.46/0.8960 \\
		MemNet     & 677K                       & 31.74/0.8893 & 28.26/0.7723 & 27.40/0.7281 & 25.50/0.7630 & 29.42/0.8942 \\
		CARN       & 1,592K                      & 32.13/0.8937 & 28.60/0.7806 & 27.58/0.7349 & 26.07/0.7837 & 30.47/0.9084 \\
		SRResNet   & 1,518K                      & {\color[HTML]{0000FE}32.17}/{\color[HTML]{FE0000}0.8951} & {\color[HTML]{FE0000}28.61}/{\color[HTML]{FE0000}0.7823} & {\color[HTML]{0000FE}27.59}/{\color[HTML]{FE0000}0.7365} & {\color[HTML]{FE0000}26.12}/{\color[HTML]{FE0000}0.7871} & {\color[HTML]{0000FE}30.48}/{\color[HTML]{0000FE}0.9087}  \\
		IMDN       & 715K                       & {\color[HTML]{FE0000}32.21}/0.8948 & 28.58/0.7811 & 27.56/0.7353 & 26.04/0.7838 & 30.45/0.9075 \\
		\textbf{PAN}  & \textbf{272K}                       & 32.13/{\color[HTML]{0000FE}0.8948} & {\color[HTML]{0000FE}28.61}/{\color[HTML]{0000FE}0.7822} & {\color[HTML]{FE0000}27.59}/{\color[HTML]{0000FE}0.7363} & {\color[HTML]{0000FE}26.11}/{\color[HTML]{0000FE}0.7854} & {\color[HTML]{FE0000}30.51}/{\color[HTML]{FE0000}0.9095} \\ \hline
		RDN        & 22,271K                     & 32.47/0.8990 & 28.81/0.7871 & 27.72/0.7419 & 26.61/0.8028 & 31.00/0.9151 \\
		RCAN       & 15,592K                     & 32.63/0.9002 & 28.87/0.7889 & 27.77/0.7436 & 26.82/0.8087 & 31.22/0.9173 \\ 
		SAN       & 15,822K                    & 32.64/0.9003 &28.92/0.7888& 27.78/0.7436 &26.79/0.8068 &31.18/0.9169 \\
		\hline
	\end{tabular}
\end{table}

  \begin{figure}[htbp]
	\centering
	\includegraphics[width=\linewidth]{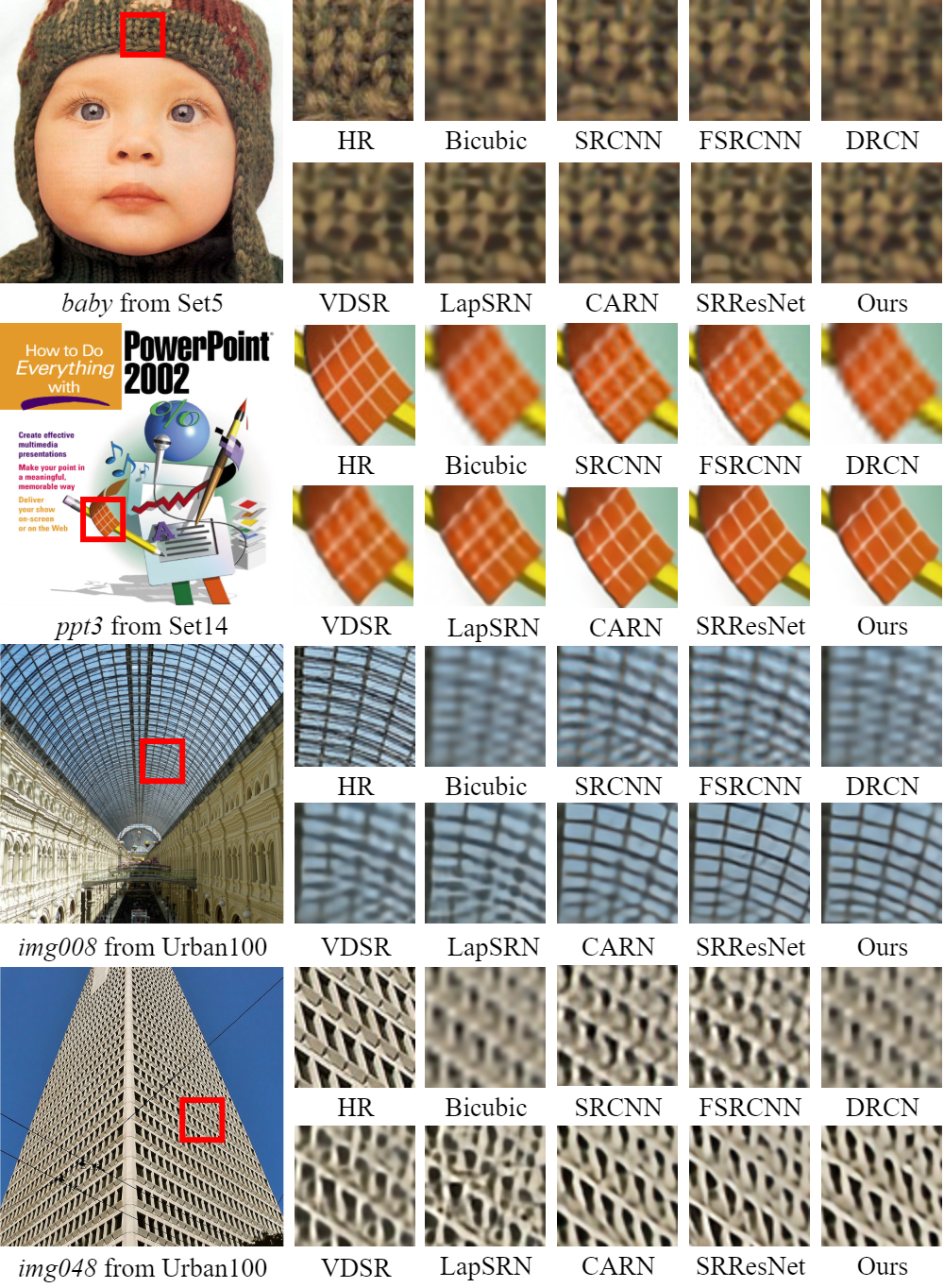}
	\caption{Visual comparison for upscaling factor $\times4$.}
	\label{fig:3}
\end{figure}

  \setlength{\tabcolsep}{4pt}
  \renewcommand{\arraystretch}{1.2}
  \begin{table}[htbp]
    \begin{center}
      \caption{The results of adding PA in different networks. We use the mean values of PSNR obtained on five datasets for upscaling factor $\times$4. We record the results in $1 \times 10^{6}$ iterations.}
      \label{tab:6}
      \begin{tabular}{llll}
        \hline
        Model       & Params & Mult-Adds                    & PSNR  \\ \hline
        PAN-SC-U    & 264K   & 27.1G                        & 28.11dB \\
        PAN         & 272K   & 28.2G                        & 28.22dB({\color[HTML]{FE0000}+0.11dB}) \\ \hline
        SRResNet    & 1,518K  & 146.1G                       & 28.22dB \\
        SRResNet-PA & 1,584K  & 149.9G                       & 28.20dB({\color[HTML]{00FF00}$-$0.02dB}) \\ \hline
        RCAN        & 15,592K & 916.9G                       & 28.75dB \\
        RCAN-PA     & 16,308K & 964.1G                       & 28.69dB({\color[HTML]{00FF00}$-$0.06dB}) \\ \hline
  
        \end{tabular}
    \end{center}
    \end{table}
  \setlength{\tabcolsep}{1.4pt}

 \subsection{Comparison with State-of-the-art Methods}
We compare the proposed PAN with commonly used lightweight SR models for upscaling factor $\times$2, $\times$3, and $\times$4, including SRCNN \cite{dong2015image}, FSRCNN \cite{dong2016accelerating}, VDSR \cite{kim2016accurate}, DRCN \cite{kim2016deeply}, LapSRN \cite{lai2017deep}, DRRN \cite{tai2017image}, MemNet \cite{tai2017memnet}, CARN \cite{ahn2018fast}, SRResNet \cite{ledig2017photo} and IMDN \cite{hui2019lightweight}. We have also listed the performance of state-of-the-art large SR models -- RDN \cite{zhang2018residual}, RCAN \cite{zhang2018image}, and SAN \cite{dai2019second} for reference. 

Table~\ref{tab:4} shows quantitative results in terms of PSNR and SSIM on 5 benchmark datasets obtained by different algorithms. In addition, the number of parameters of compared models is also given. 
From Table~\ref{tab:4}, we find that our PAN only has less than 300K parameters but outperforms most of the state-of-the-art methods. 
Specifically, CARN \cite{ahn2018fast} achieves similar performance as us, but its parameters are close to 1,592K which is about six times of ours. Compared with the baseline -- SRResNet, we could achieve higher PSNR on Set14 and Manga109 datasets. We also compare with IMDN, which is the first place of AIM 2019 Challenge on Constrained Super-Resolution and has 715K parameters. It turns out that our proposed PAN outperforms IMDN in terms of PSNR on Set14, B100 and Urban100 datasets.   

As for visual comparison, our model is able to reconstruct stripes and line patterns more accurately. For image ``ppt3", we observe that most of the compared methods generate noticeable artifacts and blurry effects while our method produces more accurate lines. For the details of the buildings in ``img008" and ``img048", PAN could achieve reconstruction with less artifacts.

\section{Conclusions}
In this work, a lightweight convolutional neural network is proposed to achieve image super resolution. In particular, we design a new pixel attention scheme, pixel attention (PA), which contains very few parameters but helps generate better reconstruction results. Besides, two building blocks for the main and reconstruction branches are proposed based on the pixel attention scheme. The SC-PA block for main branch shares similar structure as the Self-Calibrated convolution, while the U-PA block for reconstruction branch adopts the nearest-neighbor upsampling and convolution layers. This framework could improve the SR performance with little parameter cost. Experiments have demonstrated that our final model — PAN could achieve comparable performance with state-of-the-art lightweight networks.


\par\vfill\par

\clearpage
%
%
\bibliographystyle{splncs04}
\bibliography{egbib}
\end{document}